\documentclass[twocolumn,letterpaper,aps,prl,longbibliography,superscriptaddress,floatfix]{revtex4-1}

\usepackage{graphicx}
\usepackage{dcolumn}
\usepackage{bm}
\usepackage{xspace}     

\newcommand{\pt}{\mbox{$p_T$}\xspace}
\newcommand{\sqs}{\mbox{$\sqrt{s}$}\xspace}

\newcommand{\sqsntwo}{\mbox{$\sqrt{s_{_{NN}}}=200$~GeV}\xspace}
\newcommand{\sqstwo}{\mbox{$\sqrt{s}=200$~GeV}\xspace}
\newcommand{\pp}{\mbox{$p$$+$$p$}\xspace}
\newcommand{\pa}{\mbox{$p$$+$A}\xspace}
\newcommand{\pau}{\mbox{$p$$+$Au}\xspace}
\newcommand{\pal}{\mbox{$p$$+$Al}\xspace}

\newcommand{\AN}{\mbox{$A_N$}\xspace}
\newcommand{\ANB}{\mbox{$A_N^{\rm B}$}\xspace}  
\newcommand{\ANS}{\mbox{$A_N^{\rm S}$}\xspace}  

\begin{document}


\title{Nuclear dependence of the transverse-single-spin asymmetry for 
forward neutron production in polarized $p$$+$$A$ collisions at 
$\sqrt{s_{_{NN}}}=200$ GeV}

\newcommand{\abilene}{Abilene Christian University, Abilene, Texas 79699, USA}
\newcommand{\augie}{Department of Physics, Augustana University, Sioux Falls, South Dakota 57197, USA}
\newcommand{\banaras}{Department of Physics, Banaras Hindu University, Varanasi 221005, India}
\newcommand{\barc}{Bhabha Atomic Research Centre, Bombay 400 085, India}
\newcommand{\baruch}{Baruch College, City University of New York, New York, New York, 10010 USA}
\newcommand{\bnlcoll}{Collider-Accelerator Department, Brookhaven National Laboratory, Upton, New York 11973-5000, USA}
\newcommand{\bnlphys}{Physics Department, Brookhaven National Laboratory, Upton, New York 11973-5000, USA}
\newcommand{\caucr}{University of California-Riverside, Riverside, California 92521, USA}
\newcommand{\charlesczech}{Charles University, Ovocn\'{y} trh 5, Praha 1, 116 36, Prague, Czech Republic}
\newcommand{\chonbuk}{Chonbuk National University, Jeonju, 561-756, Korea}
\newcommand{\cns}{Center for Nuclear Study, Graduate School of Science, University of Tokyo, 7-3-1 Hongo, Bunkyo, Tokyo 113-0033, Japan}
\newcommand{\colorado}{University of Colorado, Boulder, Colorado 80309, USA}
\newcommand{\columbia}{Columbia University, New York, New York 10027 and Nevis Laboratories, Irvington, New York 10533, USA}
\newcommand{\czechtech}{Czech Technical University, Zikova 4, 166 36 Prague 6, Czech Republic}
\newcommand{\debrecen}{Debrecen University, H-4010 Debrecen, Egyetem t{\'e}r 1, Hungary}
\newcommand{\elte}{ELTE, E{\"o}tv{\"o}s Lor{\'a}nd University, H-1117 Budapest, P{\'a}zm{\'a}ny P.~s.~1/A, Hungary}
\newcommand{\eszterhazy}{Eszterh\'azy K\'aroly University, K\'aroly R\'obert Campus, H-3200 Gy\"ongy\"os, M\'atrai \'ut 36, Hungary}
\newcommand{\ewha}{Ewha Womans University, Seoul 120-750, Korea}
\newcommand{\fsu}{Florida State University, Tallahassee, Florida 32306, USA}
\newcommand{\gsu}{Georgia State University, Atlanta, Georgia 30303, USA}
\newcommand{\hiroshima}{Hiroshima University, Kagamiyama, Higashi-Hiroshima 739-8526, Japan}
\newcommand{\howard}{Department of Physics and Astronomy, Howard University, Washington, DC 20059, USA}
\newcommand{\ihepprot}{IHEP Protvino, State Research Center of Russian Federation, Institute for High Energy Physics, Protvino, 142281, Russia}
\newcommand{\illuiuc}{University of Illinois at Urbana-Champaign, Urbana, Illinois 61801, USA}
\newcommand{\inrras}{Institute for Nuclear Research of the Russian Academy of Sciences, prospekt 60-letiya Oktyabrya 7a, Moscow 117312, Russia}
\newcommand{\instpasczech}{Institute of Physics, Academy of Sciences of the Czech Republic, Na Slovance 2, 182 21 Prague 8, Czech Republic}
\newcommand{\isu}{Iowa State University, Ames, Iowa 50011, USA}
\newcommand{\jaea}{Advanced Science Research Center, Japan Atomic Energy Agency, 2-4 Shirakata Shirane, Tokai-mura, Naka-gun, Ibaraki-ken 319-1195, Japan}
\newcommand{\jyvaskyla}{Helsinki Institute of Physics and University of Jyv{\"a}skyl{\"a}, P.O.Box 35, FI-40014 Jyv{\"a}skyl{\"a}, Finland}
\newcommand{\kek}{KEK, High Energy Accelerator Research Organization, Tsukuba, Ibaraki 305-0801, Japan}
\newcommand{\korea}{Korea University, Seoul, 136-701, Korea}
\newcommand{\kurchatov}{National Research Center ``Kurchatov Institute", Moscow, 123098 Russia}
\newcommand{\kyoto}{Kyoto University, Kyoto 606-8502, Japan}
\newcommand{\lawllnl}{Lawrence Livermore National Laboratory, Livermore, California 94550, USA}
\newcommand{\losalamos}{Los Alamos National Laboratory, Los Alamos, New Mexico 87545, USA}
\newcommand{\lund}{Department of Physics, Lund University, Box 118, SE-221 00 Lund, Sweden}
\newcommand{\lyon}{IPNL, CNRS/IN2P3, Univ Lyon, Université Lyon 1, F-69622, Villeurbanne, France}
\newcommand{\maryland}{University of Maryland, College Park, Maryland 20742, USA}
\newcommand{\mass}{Department of Physics, University of Massachusetts, Amherst, Massachusetts 01003-9337, USA}
\newcommand{\michigan}{Department of Physics, University of Michigan, Ann Arbor, Michigan 48109-1040, USA}
\newcommand{\muhlenberg}{Muhlenberg College, Allentown, Pennsylvania 18104-5586, USA}
\newcommand{\nara}{Nara Women's University, Kita-uoya Nishi-machi Nara 630-8506, Japan}
\newcommand{\natmephi}{National Research Nuclear University, MEPhI, Moscow Engineering Physics Institute, Moscow, 115409, Russia}
\newcommand{\newmex}{University of New Mexico, Albuquerque, New Mexico 87131, USA}
\newcommand{\nmsu}{New Mexico State University, Las Cruces, New Mexico 88003, USA}
\newcommand{\ohio}{Department of Physics and Astronomy, Ohio University, Athens, Ohio 45701, USA}
\newcommand{\ornl}{Oak Ridge National Laboratory, Oak Ridge, Tennessee 37831, USA}
\newcommand{\orsay}{IPN-Orsay, Univ.~Paris-Sud, CNRS/IN2P3, Universit\'e Paris-Saclay, BP1, F-91406, Orsay, France}
\newcommand{\peking}{Peking University, Beijing 100871, People's Republic of China}
\newcommand{\pnpi}{PNPI, Petersburg Nuclear Physics Institute, Gatchina, Leningrad region, 188300, Russia}
\newcommand{\riken}{RIKEN Nishina Center for Accelerator-Based Science, Wako, Saitama 351-0198, Japan}
\newcommand{\rikjrbrc}{RIKEN BNL Research Center, Brookhaven National Laboratory, Upton, New York 11973-5000, USA}
\newcommand{\rikkyo}{Physics Department, Rikkyo University, 3-34-1 Nishi-Ikebukuro, Toshima, Tokyo 171-8501, Japan}
\newcommand{\saispbstu}{Saint Petersburg State Polytechnic University, St.~Petersburg, 195251 Russia}
\newcommand{\seoulnat}{Department of Physics and Astronomy, Seoul National University, Seoul 151-742, Korea}
\newcommand{\stonybrkc}{Chemistry Department, Stony Brook University, SUNY, Stony Brook, New York 11794-3400, USA}
\newcommand{\stonycrkp}{Department of Physics and Astronomy, Stony Brook University, SUNY, Stony Brook, New York 11794-3800, USA}
\newcommand{\tenn}{University of Tennessee, Knoxville, Tennessee 37996, USA}
\newcommand{\titech}{Department of Physics, Tokyo Institute of Technology, Oh-okayama, Meguro, Tokyo 152-8551, Japan}
\newcommand{\tsukuba}{Center for Integrated Research in Fundamental Science and Engineering, University of Tsukuba, Tsukuba, Ibaraki 305, Japan}
\newcommand{\vandy}{Vanderbilt University, Nashville, Tennessee 37235, USA}
\newcommand{\weizmann}{Weizmann Institute, Rehovot 76100, Israel}
\newcommand{\wigner}{Institute for Particle and Nuclear Physics, Wigner Research Centre for Physics, Hungarian Academy of Sciences (Wigner RCP, RMKI) H-1525 Budapest 114, POBox 49, Budapest, Hungary}
\newcommand{\yonsei}{Yonsei University, IPAP, Seoul 120-749, Korea}
\newcommand{\zagreb}{Department of Physics, Faculty of Science, University of Zagreb, Bijeni\v{c}ka c.~32 HR-10002 Zagreb, Croatia}
\affiliation{\abilene}
\affiliation{\augie}
\affiliation{\banaras}
\affiliation{\barc}
\affiliation{\baruch}
\affiliation{\bnlcoll}
\affiliation{\bnlphys}
\affiliation{\caucr}
\affiliation{\charlesczech}
\affiliation{\chonbuk}
\affiliation{\cns}
\affiliation{\colorado}
\affiliation{\columbia}
\affiliation{\czechtech}
\affiliation{\debrecen}
\affiliation{\elte}
\affiliation{\eszterhazy}
\affiliation{\ewha}
\affiliation{\fsu}
\affiliation{\gsu}
\affiliation{\hiroshima}
\affiliation{\howard}
\affiliation{\ihepprot}
\affiliation{\illuiuc}
\affiliation{\inrras}
\affiliation{\instpasczech}
\affiliation{\isu}
\affiliation{\jaea}
\affiliation{\jyvaskyla}
\affiliation{\kek}
\affiliation{\korea}
\affiliation{\kurchatov}
\affiliation{\kyoto}
\affiliation{\lawllnl}
\affiliation{\losalamos}
\affiliation{\lund}
\affiliation{\lyon}
\affiliation{\maryland}
\affiliation{\mass}
\affiliation{\michigan}
\affiliation{\muhlenberg}
\affiliation{\nara}
\affiliation{\natmephi}
\affiliation{\newmex}
\affiliation{\nmsu}
\affiliation{\ohio}
\affiliation{\ornl}
\affiliation{\orsay}
\affiliation{\peking}
\affiliation{\pnpi}
\affiliation{\riken}
\affiliation{\rikjrbrc}
\affiliation{\rikkyo}
\affiliation{\saispbstu}
\affiliation{\seoulnat}
\affiliation{\stonybrkc}
\affiliation{\stonycrkp}
\affiliation{\tenn}
\affiliation{\titech}
\affiliation{\tsukuba}
\affiliation{\vandy}
\affiliation{\weizmann}
\affiliation{\wigner}
\affiliation{\yonsei}
\affiliation{\zagreb}
\author{C.~Aidala} \affiliation{\michigan} 
\author{Y.~Akiba} \email[PHENIX Spokesperson: ]{akiba@rcf.rhic.bnl.gov} \affiliation{\riken} \affiliation{\rikjrbrc} 
\author{M.~Alfred} \affiliation{\howard} 
\author{V.~Andrieux} \affiliation{\michigan} 
\author{K.~Aoki} \affiliation{\kek} 
\author{N.~Apadula} \affiliation{\isu} 
\author{H.~Asano} \affiliation{\kyoto} \affiliation{\riken} 
\author{C.~Ayuso} \affiliation{\michigan} 
\author{B.~Azmoun} \affiliation{\bnlphys} 
\author{V.~Babintsev} \affiliation{\ihepprot} 
% ...
\collaboration{PHENIX Collaboration} \noaffiliation

\date{\today}


\begin{abstract}

During 2015 the Relativistic Heavy Ion Collider (RHIC) provided 
collisions of transversely polarized protons with Au and Al nuclei for 
the first time, enabling the exploration of transverse-single-spin 
asymmetries with heavy nuclei.  Large single-spin asymmetries in very 
forward neutron production have been previously observed in transversely 
polarized $p$$+$$p$ collisions at RHIC, and the existing theoretical 
framework that was successful in describing the single-spin asymmetry in 
$p$$+$$p$ collisions predicts only a moderate atomic-mass-number ($A$) 
dependence.  In contrast, the asymmetries observed at RHIC in $p$$+$$A$ 
collisions showed a surprisingly strong $A$ dependence in inclusive 
forward neutron production. The observed asymmetry in $p$$+$Al 
collisions is much smaller, while the asymmetry in $p$$+$Au collisions 
is a factor of three larger in absolute value and of opposite sign.  The 
interplay of different neutron production mechanisms is discussed as a 
possible explanation of the observed $A$ dependence.

\end{abstract}

\maketitle



Understanding forward particle production in high energy hadron 
collisions is of great importance, because most of the energy goes  
in the forward direction, and therefore informs our understanding of 
overall particle production.  
This has particular importance in studies 
of ultra-high energy cosmic rays, 
where extraction of the cosmic ray distributions from air shower measurements 
depends on models of forward particle production in the interaction with nuclei 
in the air~\cite{dEnterria201198,Kampert2012660,lhcf_collaboration_measurements_2016}.
Mechanisms for forward 
particle production are not well understood, as perturbative quantum 
chromodynamics (pQCD) is not applicable at small momentum transfers and 
diffractive production mechanisms are not well modeled. To better 
understand production mechanisms, measurement of the single spin 
asymmetry $A_N$, describing the azimuthal asymmetry of particle 
production relative to the spin direction of the transversely polarized 
beam or target provides crucial tests and deeper insight beyond just 
cross-section measurements. The spin degree of freedom has 
served as a strong discriminator between theoretical models.  For 
example, the origin of the large asymmetries discovered in forward meson 
production in \pp collisions from 
\sqs=4.9--19.4~GeV~\cite{SSA1,SSA2,SSA3,SSA4,E925,E704ch,E704pi0,E704eta} 
and later confirmed at \sqs=62.4--500 GeV at the Relativistic Heavy Ion 
Collider 
(RHIC)~\cite{BRAHMS,STARpi0,STAReta,PHENIX_MPC,STAR500,phenix_eta_an} 
has been under intensive discussion for three decades and still remains 
an open question~\cite{review}.  Despite substantial theoretical attempts 
to reproduce data in the pQCD regime using the conventional 
$2\rightarrow2$ parton scattering processes, the latest multiplicity 
dependent $A_N$ measurements from RHIC~\cite{STAR-photon} indicate that 
a significant contribution to the asymmetry may be of a diffractive 
nature.

Another important approach in forward particle production is 
to study the nuclear dependence in \pa collisions. 
In the perturbative region, theoretical approaches based on 
color-glass-condensate models predicted that hadronic $A_N$ should 
decrease with increasing $A$~\cite{boer_single_2006,boer_polarized_2003,boer_saturation_2009,kang_single_2011,kovchegov_new_2012}, 
while some approaches 
based on pQCD factorization predicted that $A_N$ would stay approximately the same 
for all nuclear targets~\cite{Qiu_2012}. 
On the other hand, almost no theoretical/experimental studies are available 
in the nonperturbative region or diffractive scattering with polarized probes on nuclei, 
and interesting phenomena may be hidden in this unexplored region.

In the case of forward neutron production in $p$$+$$p$ collisions, production 
cross sections~\cite{ISR1,ISR2,IP12_Neutron} were successfully explained 
in terms of one-pion 
exchange~\cite{Capella75,Boris96,Nikolaev,Kaidalov,Boris_ANpT}. However, 
that model could not explain the sizable $A_N$ in very forward (near 
zero degree) neutron production, discovered at RHIC in \pp collisions at 
\sqstwo~\cite{IP12_Neutron}.  To reproduce the experimental asymmetry, 
an interference between the spin-flip $\pi$ exchange and a non spin-flip 
$a_1$-Reggeon exchange was necessary~\cite{Boris_ANpT}.  Kopeliovich, 
Potashnikova, and Schmidt considered nuclear absorption effects as a 
source for a possible $A$ dependence of $A_N$, and found only a small 
effect~\cite{Boris2016}.

In this Letter, we report the first measurements of $A_N$ for very 
forward neutron production in collisions between polarized protons 
and nuclei (Al and Au) at \sqsntwo recorded in 2015 with the PHENIX 
detector~\cite{PHENIX-detector}.  For \pp collisions 18 
RHIC stores were used and 1 store each for \pal and \pau 
measurements, with a typical store length of 8 hours. The average 
beam polarization in \pp, \pal, and \pau data samples was $0.515 
\pm 0.002$, $0.59 \pm 0.02$ and $0.59 \pm 0.04$, respectively, with 
additional global uncertainty of 3\% from the polarization 
normalization~\cite{schmidke2013rhic,schmidkeemail}.

\begin{figure}[htb!]
\begin{center}
\includegraphics[width=1.0\linewidth]{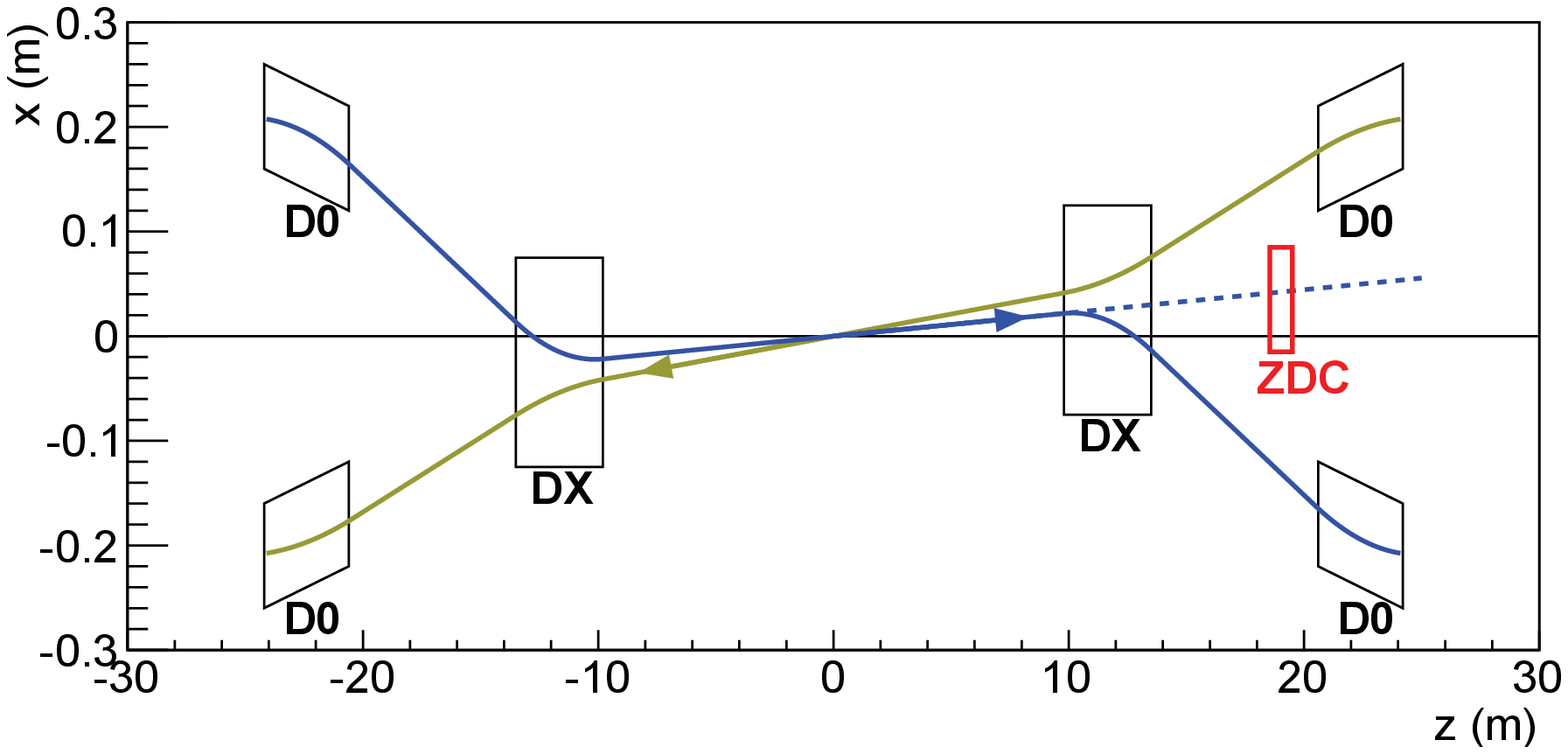}
\end{center}
\caption{\label{fig:paFig}
ZDC location and beam orbits of proton (blue) beam and heavy-ion 
(yellow) beam in the special stores used for this analysis; the $z$-axis 
shows the nominal beam direction, and the dashed line represents the 
zero-degree neutron trajectory. DX and D0 are the RHIC beam bending 
dipole magnets.}
\end{figure}

The experimental setup using a zero-degree calorimeter 
(ZDC)~\cite{Adler2001488} and a position-sensitive shower-maximum 
detector (SMD) is similar to the one used for \pp 
data~\cite{PHENIX_Neutron}.  The ZDC comprises three modules located in 
series at $\pm$18 m away from the collision point.  The ZDC has an 
acceptance in the transverse plane of 10 $\times$ 10 cm$^2$, with a 
total of 5.1 nuclear interaction lengths (or 149 radiation lengths), and 
an energy resolution of $\sim$25\%--20\% for 50--100 GeV neutrons. The SMD 
comprises $x$-$y$ (horizontal-vertical) scintillator strip hodoscopes 
inserted between the first and second ZDC modules (approximately at the 
position of the maximum hadronic shower), and provides a position 
resolution of $\sim 1$ cm for 50--100 GeV neutrons.  These detectors are 
located downstream of the RHIC DX beam splitting magnet, so that near 
beam-momentum charged particles from collisions are expected to be swept 
into the beam lines and out of the ZDC acceptance 
(see Fig.~\ref{fig:paFig}).

To accommodate asymmetric \pa collisions of beams with different 
rigidity, the DX magnets were moved horizontally~\cite{BeamOptics}. In 
this special setup for the present measurement, the proton beam was 
angled off axis by ${\sim}2$ mrad relative to the nominal beam direction 
at the collision point, with a crossing angle with the Au (Al) beam of 
2.0 mrad (1.1 mrad).  Correspondingly, the ZDC was moved by 3.6 cm 
(2 mrad) to keep zero-degree neutrons at the ZDC center 
(see Fig.~\ref{fig:paFig}).

The data was collected with triggers employing the ZDC and beam-beam 
counters (BBCs)~\cite{Allen2003549}. Only the north ZDC detector, facing 
the incoming polarized proton beam was used in this analysis. Two BBC 
counters are located at $\pm144$ cm from the nominal collision point 
along the beam pipe and are designed to detect charged particles in the 
pseudorapidity range of $\pm$(3.0--3.9) with full azimuthal coverage. The 
ZDC inclusive trigger required the energy deposited in the ZDC to be 
greater than 15 GeV. The ZDC$\otimes$BBC-tag trigger in addition 
required at least one hit in each of the BBCs, and ZDC$\otimes$BBC-veto 
trigger required no hits in both BBCs. The latter two sets represent 
mutually exclusive but not complete subsets of the ZDC inclusive 
triggered data.

As described in detail in Ref.~\cite{PHENIX_Neutron}, event selection 
and neutron identification cuts include: (1) a total ZDC energy cut of 
40--120 GeV; (2) at least two SMD strips fired (above threshold) in both 
$x$ and $y$ directions, and a nonzero (above threshold) energy in the 
second ZDC module (to reject photons); and (3) an acceptance cut of 0.5 
$<r<$ 4.0 cm for the reconstructed radial distance $r$ from the 
determined beam center (to reduce the impact of the position resolution 
and edge effects in the asymmetry measurements).

The raw asymmetry ($\epsilon_{N}(\phi)$) is calculated using the 
square-root formula~\cite{PHENIX_Neutron} for each azimuthal angle 
($\phi$) bin. The polarization normalized $A_N^{\rm fit}$ is then 
extracted from the fit to a sine function
\begin{equation}
  \epsilon_{N}(\phi)  = P A_{N}^{\rm fit} \sin{(\phi-\phi_{0})},
  \label{eq:meas_asym}
\end{equation}
where $P$ is the proton beam polarization and $\phi_{0}$ is the 
polarization direction in the transverse plane.

\begin{figure}[htb!]
\centering
\includegraphics[width=1.0\linewidth]{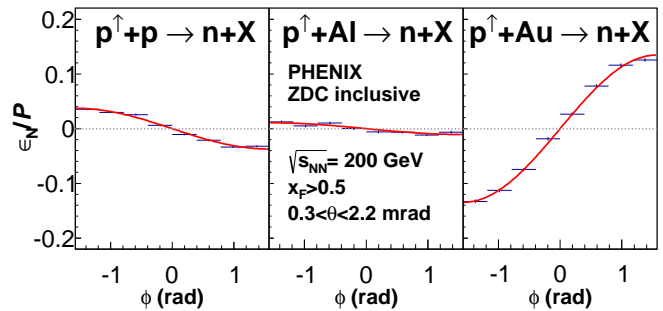}
\caption{\label{fig:epsilon_phi}$A_{N}^{\rm fit}$ 
fit of ZDC inclusive samples.}
\end{figure}

Figure~\ref{fig:epsilon_phi} compares $\epsilon_N(\phi)/P$ results for 
ZDC inclusive samples from \pp, \pal and \pau collisions and shows the 
nuclear dependence of $A_N^{\rm fit}$, including a sign change from 
negative in \pp collisions to positive in \pau collisions. 
The $A_N^{\rm fit}$ was measured separately in each PHENIX data taking 
segment, typically 60 min long, and then the weighted average was calculated.
The obtained $A_N^{\rm fit}$ is then corrected for backgrounds and detector 
responses. The main background contribution comes from protons, generated 
by elastic, diffractive, and hard processes.

Protons from elastic and diffractive reactions travel close to the beam 
line and are swept by the DX magnet to the right (toward negative $x$ in 
Fig.~\ref{fig:paFig}). Only a small fraction of such protons scattered 
by large angles, larger than 4--5 mrad, fall in the ZDC acceptance. 
Because the cross section for these reactions falls sharply with 
scattering angle, these protons contribute mainly on the right side of 
the ZDC. This contribution was evaluated from the particle position 
distribution as measured by the SMD and found to be 9\% and 32\% in the 
inclusive ZDC and ZDC$\otimes$BBC-veto triggered samples respectively in 
\pp collisions, $<2\%$ in both samples in \pa collisions, and negligible 
in ZDC$\otimes$BBC-tag samples of both \pp and \pa collisions. The 
significant suppression of elastic and diffractive proton background 
relative to the neutron signal in \pa collisions can be understood as 
due to the stronger magnetic fields in the DX magnets. Correspondingly, 
the minimum scattering angle for the elastic and diffractive proton 
backgrounds to reach the ZDC acceptance increases from 3.8 mrad to 5 
mrad, leading to a cross section reduction by an order of magnitude.

The contribution of charged hadron background from hard scattering 
processes, distributed nearly uniformly over the ZDC acceptance, was 
estimated using {\sc pythia}{\footnotesize 6}~\cite{SJOSTRAND2001238} 
with a {\sc geant}{\footnotesize 3}~\cite{GEANT} detector simulation.  
However, from previous studies where a charge veto counter was 
installed in front of the ZDC to measure the charged hadron background, 
it was found that simulation underestimates the proton background by a 
factor of $\sim$2~\cite{PHENIX_Neutron}. Therefore the hard scattering 
background contribution from simulation was scaled by a factor of two 
with an uncertainty equal to the size of the increase.  In \pp 
collisions this background fraction resulted in 6$\pm$3\%, 
3$\pm$1.5\% and 12$\pm$6\% in ZDC, ZDC$\otimes$BBC-veto and 
ZDC$\otimes$BBC-tag triggered samples, respectively. In \pa collisions 
due to increased neutron signal from electromagnetic (EM) processes (to 
be discussed later), the relative background contributions are expected 
to be smaller. Therefore the measured asymmetries in \pa collisions 
were not corrected for background, but one-sided systematic 
uncertainties (in the direction of asymmetry magnitude increase) equal to 
the upper $1\sigma$ limit of the background fractions 
taken from the $p$$+$$p$ case, i.e. 
9\%, 4.5\% and 18\%, were conservatively assigned in ZDC, 
ZDC$\otimes$BBC-veto and ZDC$\otimes$BBC-tag triggered samples, 
respectively.

From the considerations above, only the \pp asymmetries were corrected 
for backgrounds according to
\begin{equation}
A_N^{\rm S}=\frac{\AN^{\rm fit}-r_{\rm eff}A_N^{\rm B}}{1-r_{\rm eff}},
\end{equation}
where \ANS and \ANB stand for signal and background asymmetries, 
and $r_{\rm eff}$ is the ``effective'' background fraction in the 
reconstructed neutron sample. The parameter $r_{\rm eff}$ accounts for 
the dilution of the background effect in $A_N^{\rm fit}$ in the case 
when the background contributes preferably on one side of the detector 
(as from elastic or diffractive protons). 
This effect, which was studied in simulation, comes from a specific way
the left and right sides of detector acceptance are combined in the
square-root formula for asymmetry calculation.  The background asymmetry 
\ANB was evaluated from the comparison of asymmetries with and 
without the charge veto cut from the 2008 data when the charge veto 
counter was available, and then used in Eq.~(2). 
The asymmetries \ANB were found to be 
consistent with zero within statistical uncertainties for all triggers.
After background correction, \ANS results for $\pp$ from 2008 and 
2015 data were found to be consistent within statistical uncertainties. 
Asymmetries from 2015 data were used in the final results.

Besides charged hadrons, the other background sources are photons and 
$K^0$ mesons. From {\sc pythia}{\footnotesize 6} simulation their 
contribution after the analysis cuts was evaluated to be below 3\% in 
all collision systems and triggers, and was neglected in the asymmetry 
results.

The measured asymmetries are affected by detector resolutions and other 
detector systematic effects (e.g. edge effects), as well as by the 
uncertainty in the shape of the neutron production cross section vs \pt 
and $x_F$, the size of the asymmetry, 
and the assumption for the shape of $A_N(p_T)$ within the 
\pt range sampled in this analysis. These effects were studied in 
detail with a {\sc geant}{\footnotesize 3} Monte Carlo simulation. The 
fully corrected transverse single spin asymmetry $A_N$ was calculated 
as $A_{N} = A_{N}^{S}/C_{\phi}$ where the correction factor $C_{\phi}$ 
was calculated in the simulation as the ratio of the measured asymmetry 
to the average input asymmetry over the neutron sample collected with 
experimental cuts used in the analysis. The biggest variation in 
$C_{\phi}$ comes from the position resolution uncertainty and the 
assumption for $A_N(p_T)$. The position resolution in simulation vs data 
was confirmed from the comparison of shower shape and its fluctuations 
in SMD strips. The simulation was tuned to data by varying noise and 
thresholds in the SMD channels, as 
well as by introducing a cross talk effect, similar 
to~\cite{PHENIX_Neutron}. An overall value of 3\% was assigned to the 
$C_{\phi}$ uncertainty. For the shape of $A_N(p_T)$, it was modeled as 
$A_N(p_T)={\rm const}$ (as was assumed in~\cite{PHENIX_Neutron}) and 
$A_N(p_T) \propto p_T$ (which is supported by theory 
in the $p_T$ range relevant here~\cite{Boris_ANpT}). 
The difference of 3\% was included in the 
$C_{\phi}$ uncertainty. The final correction factor applied to the 
measured asymmetries is $C_{\phi}=0.855 \pm 0.036$.
Note, the $C_{\phi}$ value here is higher than the one in our previous 
publication~\cite{PHENIX_Neutron} mainly due to two reasons: 
first, more realistic $A_N(p_T) \propto p_T$ assumption was used in this 
analysis, and, second, the optimized SMD thresholds reduced the smearing 
effect.

In addition to the beam polarization, 
background, and smearing correction ($C_{\phi}$) discussed above, the 
other sources of systematic uncertainties are the ZDC and SMD gain 
calibrations (including threshold variation) and location of 
the beam center on the ZDC plane.
The latter is among the dominant uncertainties in this data, 
contributing 0.002--0.010 to the $A_N$ uncertainty. 
It was estimated by calculating the asymmetry for varying assumptions 
of the beam axis projection on the ZDC plane, $\pm 1$ cm in horizontal 
and $\pm 0.5$ cm in vertical directions from the ZDC center, 
which reflect the uncertainty in ZDC alignment relative to the beam axis. 

\begin{figure}[htb!]
\includegraphics[width=1.0\linewidth]{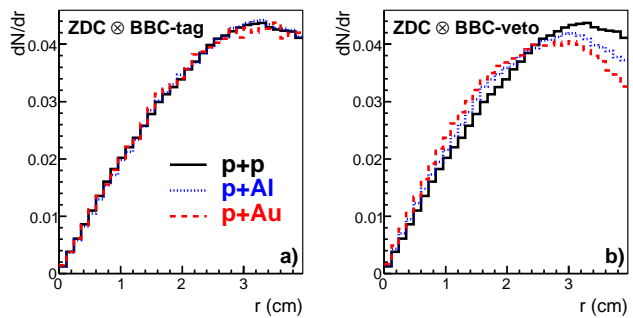}
\caption{\label{fig:r}
The $r$ distribution of (a) ZDC$\otimes$BBC-tag sample and 
(b) ZDC$\otimes$BBC-veto sample for three collision systems.}
\end{figure}

The analyzed data correspond to the neutron sampled \pt in the range 
smaller than $0.25$ GeV/$c$ peaked at about 0.1 GeV/$c$, which is 
defined mainly by detector acceptance and which is
affected by detector resolutions. Due to the varying contribution of 
different processes to neutron production, the sampled \pt distribution 
may vary in different collision systems and in different triggered data. 
Figure~\ref{fig:r} shows the differences in the radial distributions, 
which is related to the neutron production cross section $d\sigma/dp_T$ 
by $p_T \propto r$~\cite{PHENIX_Neutron}. From a comparison with 
simulation assuming different slope parameter, $b$, in the 
parameterization $d\sigma/dp_T \sim e^{-b \cdot p_T}$, the data were 
found to be consistent with $b=4$ (GeV/$c$)$^{-1}$ for all collision 
systems in ZDC$\otimes$BBC-tag triggered data, and $b=$4, 6 and 8 
(GeV/$c$)$^{-1}$ in \pp, \pal and \pau collisions, respectively, in 
ZDC$\otimes$BBC-veto triggered sample, with uncertainty $\sigma_b=1$ 
(GeV/$c$)$^{-1}$ reflecting its sensitivity to SMD gain calibration and 
thresholds. These variations lead to a difference in the average \pt 
sampled in different collision systems and triggers by as much as 10\%.
As can be also judged from Fig.~\ref{fig:r}, due to the small detector 
acceptance, the sampled \pt distribution shows very modest dependence on 
the slope of the input \pt distribution, particularly at low \pt (or $r$), 
which is most responsible for the dilution of the measured 
asymmetry.  As a consequence,  the variation of the correction factor 
$C_{\phi}$ due to different slope parameters $b$ discussed above was 
less than 1\%.

\begin{figure}[htb!]
\centering
\includegraphics[width=1.0\linewidth]{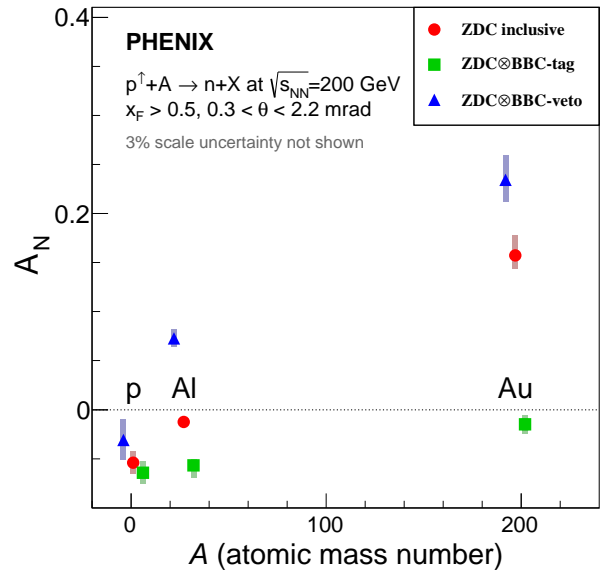}
\caption{\label{fig:Run15_AN_Result}Forward neutron $A_N$ in \pa 
collisions for $A=$1 ($p$), 27 (Al) and 197 (Au), 
for ZDC inclusive, ZDC$\otimes$BBC-tag and ZDC$\otimes$BBC-veto 
triggered samples; color bars are systematic uncertainties, 
statistical uncertainties are smaller than the marker size;
3\% scale uncertainty (not shown) is from 
polarization normalization uncertainty.
Data points are shifted horizontally for better visibility.
}
\end{figure}

Figure~\ref{fig:Run15_AN_Result} and Table~\ref{t:syst} summarize the 
results for $A_N$ in forward neutron production in \pp, \pal and \pau 
collisions, for ZDC inclusive, ZDC$\otimes$BBC-tag and 
ZDC$\otimes$BBC-veto samples. 
In addition to 3\% scale uncertainty from polarization normalization, 
common to all points, the other part of the polarization uncertainty 
is correlated for different triggers in a particular collision system.
The presented asymmetries in \pp collisions are consistent with our previous 
publication~\cite{PHENIX_Neutron}, albeit with larger systematic 
uncertainties in this data due to larger background 
(unlike this measurement, charged veto counter was used 
in~\cite{PHENIX_Neutron} to suppress the background), and larger variations 
due to uncertainty of the beam position on the ZDC plane.

\setlength{\tabcolsep}{4pt}
\begingroup \squeezetable
\begin{table*}
\caption{\label{t:syst}
$A_N$ for forward neutron production in \pp, \pal, and \pau collisions, 
for ZDC inclusive, ZDC$\otimes$BBC-tag, and ZDC$\otimes$BBC-veto 
samples.}
    \begin{ruledtabular}  \begin{tabular}{ l c c c c c c c c c c c c }
&& & \pp &  && & \pal  & && & \pau &   \\   
&& Inclusive& BBC Tag& BBC Veto
&& Inclusive& BBC Tag& BBC Veto
&& Inclusive& BBC Tag& BBC Veto  \\  \hline 
$A_N$  && -0.054  & -0.064 & -0.031 
&& -0.013 & -0.057 & 0.073 
&& 0.157  & -0.015  & 0.234 \\
Statistical error  
&&  $\pm$0.001 &  $\pm$0.002 & $\pm$0.004
&& $\pm$0.002 &  $\pm$0.003 & $\pm$0.003 
&&  $\pm$0.002  & $\pm$0.005 & $\pm$0.002 \\ 
Systematic error: &&&&&&&&&&&& \\
 \it{   Background}   
&& $\pm$0.007 & $\pm$0.009 & $\pm$0.017 
&& -0.001     & -0.010     & +0.004     
&& +0.015     & -0.003     & +0.012       \\   
 \it{   Smearing}     
&& $\pm$0.002 & $\pm$0.003 & $\pm$0.001 
&& $< 0.001$ & $\pm$0.002 & $\pm$0.003 
&& $\pm$0.007 & $< 0.001$ & $\pm$0.010   \\   
 \it{   Beam pos.}    
&& $\pm$0.009 & $\pm$0.006 & $\pm$0.010 
&& $\pm$0.004 & $\pm$0.004 & $\pm$0.006 
&& $\pm$0.002 & $\pm$0.004 & $\pm$0.008   \\  
 \it{   Polarization} 
&& $< 0.001$ & $< 0.001$ & $< 0.001$ 
&& $< 0.001$ & $\pm$0.002 & $\pm$0.003 
&& $\pm$0.011 & $\pm$0.001 & $\pm$0.017   \\  
 \it{   Calibration}  
&& $\pm$0.003 & $\pm$0.001 & $\pm$0.007 
&& $\pm$0.001 & $\pm$0.004 & $\pm$0.004 
&& $\pm$0.004 & $\pm$0.009 & $\pm$0.006   \\   
 \it{   Total systematic}   
&& $\pm$0.012 & $\pm$0.011 & $\pm$0.021 
&& $^{+0.004}_{-0.004}$ & $^{+0.007}_{-0.012}$ & $^{+0.009}_{-0.008}$ 
&& $^{+0.020}_{-0.014}$ & $^{+0.009}_{-0.010}$ & $^{+0.025}_{-0.022}$   \\   
  \end{tabular}      \end{ruledtabular}
\end{table*}     	
\endgroup


From Fig.~\ref{fig:Run15_AN_Result}, the $A$ dependence of $A_N$ for 
inclusive neutrons is strong. Compared to the $A_N$ of \pp collisions, 
the observed asymmetry in \pal collisions is much smaller, while the 
asymmetry in \pau collisions is a factor of three larger in absolute 
value and of opposite sign. This behavior is unexpected because the 
theoretical framework using $\pi$ and $a_1$-Reggeon interference can 
only predict a moderate nuclear dependence, and there is no known 
mechanism to flip the sign of $A_N$ within this framework~\cite{Boris2016}.

The asymmetries requiring BBC hits are remarkably different. Once BBC 
hits are required (ZDC$\otimes$BBC-tag), the drastic behavior of the 
inclusive $A_{N}$ vanishes and its sign stays negative, approaching 
$A_N=0$ at large \mbox{$A$}\xspace.  In contrast, the strong 
\mbox{$A$}\xspace dependence is amplified once no hits in the BBC are 
required (ZDC$\otimes$BBC-veto). While the BBCs cover a limited 
acceptance, the requirement (or veto) of hits in the BBC should place 
constraints on the activity near the detected neutron and thus the 
corresponding production mechanism.


One possibility to explain the present results is a contribution from EM 
interactions, which have been demonstrated to be important for reactions 
with small momentum transfer, e.g., in ultra-peripheral heavy ion 
collision at RHIC~\cite{upcstar08,upcphenix09,upcstar10,upcstar12} and 
Large Hadron 
Collider~\cite{upcalice13p,upcalice13e,upcalice14,upcalice15}, including 
forward neutron production in p+A collisions~\cite{UPC_MC}, and 
polarization observables in fixed target experiments~\cite{CNI,Fermi}. 
Although it was ignored in the interpretation for the \pp 
data~\cite{Boris2016}, EM interactions become increasingly important for 
large atomic number ($Z$) nuclei, as the EM field of the nucleus is a 
rich source of virtual photons, increasing as $Z^2$. Forward neutrons in 
the final state can be produced through nonresonant photo-$\pi^+$ 
production and neutron decay channel from photo-nucleon excitation 
processes, such as the $\Delta$ resonance~\cite{Mitsuka:2017czj}.

According to a Monte-Carlo study~\cite{UPC_MC}, the neutron and its associated 
$\pi^+$ produced through this process are substantially boosted towards 
the proton beam direction, so that only a small fraction of pions would 
be detected by the BBC. Thus, a large fraction of EM processes are 
expected to be suppressed in the ZDC$\otimes$BBC-tag events while 
enhanced in the ZDC$\otimes$BBC-veto events. Here, it is noted that the 
importance of EM processes in \pa collisions is also hinted at in the 
present data: the ratio between reconstructed neutrons in 
ZDC$\otimes$BBC-veto and ZDC$\otimes$BBC-tag samples increases from 
smaller than $0.5$ in \pp to $\sim1$ ($\sim5$) in \pal (\pau) 
collisions. In addition, a faster drop of the neutron production cross 
section with \pt in \pa collisions in ZDC$\otimes$BBC-veto triggered 
data discussed in Fig.~\ref{fig:r}b is consistent with increasing role 
of EM processes that have softer \pt distribution than hadronic 
processes.

Similarly in the asymmetry measurements, contributions of different 
production mechanisms may be suppressed or enhanced by different event 
selection triggers. Hence, while the result for the ZDC$\otimes$BBC-tag 
sample may be explained by the conventional pion and $a_1$-Reggeon 
interference mechanism~\cite{Boris2016}, that for the 
ZDC$\otimes$BBC-veto triggered sample could be explained by 
contributions from interference with EM amplitudes~\cite{Mitsuka:2017czj},
which are expected to 
be enhanced in that dataset.  However, there could be other 
mechanisms, such as diffractive scattering, which is also expected to be 
enhanced by a ZDC$\otimes$BBC-veto trigger.  Therefore, further studies 
are needed to fully understand the present results.

In summary, we observe an unexpectedly strong $A$ dependence in $A_N$ of 
inclusive forward neutron production in polarized \pa collisions at 
\sqsntwo.  Furthermore, a distinctly different behavior of $A_N$ was 
observed in two oppositely trigger-enhanced data sets. These surprising 
behaviors could be explained by a contribution of EM interactions, which 
may be sizable for heavy nuclei.  Further studies of the production 
mechanism including EM contributions and diffractive scattering would 
have an impact not only to hadron physics, but also to cosmic-ray 
science, where measurements of high-energy cosmic rays depend on models 
of forward particle production in the interactions with nuclei in the 
air.  Spin asymmetry measurements not only provide a unique discriminating 
power for the models of particle production, but also will contribute 
to our understanding of the origin of the transverse spin asymmetries 
in hadronic collisions.



We thank the staff of the Collider-Accelerator and Physics
Departments at Brookhaven National Laboratory, especially the
CA-D staff for providing beams with a special tune
for these measurements, and the staff of
the other PHENIX participating institutions for their vital
contributions.
We also thank Boris Kopeliovich and Michal K$\check{\rm r}$elina for 
providing us with theoretical calculations of the elastic proton cross 
sections and for useful discussions. 
We acknowledge support from the Office of Nuclear Physics in the
Office of Science of the Department of Energy,
the National Science Foundation, 
Abilene Christian University Research Council, 
Research Foundation of SUNY, and
Dean of the College of Arts and Sciences, Vanderbilt University 
(U.S.A),
Ministry of Education, Culture, Sports, Science, and Technology
and the Japan Society for the Promotion of Science (Japan),
Conselho Nacional de Desenvolvimento Cient\'{\i}fico e
Tecnol{\'o}gico and Funda\c c{\~a}o de Amparo {\`a} Pesquisa do
Estado de S{\~a}o Paulo (Brazil),
Natural Science Foundation of China (People's Republic of China),
Croatian Science Foundation and
Ministry of Science and Education (Croatia),
Ministry of Education, Youth and Sports (Czech Republic),
Centre National de la Recherche Scientifique, Commissariat
{\`a} l'{\'E}nergie Atomique, and Institut National de Physique
Nucl{\'e}aire et de Physique des Particules (France),
Bundesministerium f\"ur Bildung und Forschung, Deutscher
Akademischer Austausch Dienst, and Alexander von Humboldt Stiftung (Germany),
J. Bolyai Research Scholarship, EFOP, the New National Excellence
Program ({\'U}NKP), NKFIH, and OTKA (Hungary),
Department of Atomic Energy and Department of Science and Technology (India), 
Israel Science Foundation (Israel), 
Basic Science Research Program through NRF of the Ministry of Education (Korea),
Physics Department, Lahore University of Management Sciences (Pakistan),
Ministry of Education and Science, Russian Academy of Sciences,
Federal Agency of Atomic Energy (Russia),
VR and Wallenberg Foundation (Sweden), 
the U.S. Civilian Research and Development Foundation for the
Independent States of the Former Soviet Union, 
the Hungarian American Enterprise Scholarship Fund,
the US-Hungarian Fulbright Foundation,
and the US-Israel Binational Science Foundation.



%
 
\end{document}